\documentclass[12pt]{iopart}
\usepackage[utf8]{inputenc}
\usepackage[english]{babel}

\usepackage{amssymb}
\usepackage{graphicx}

\graphicspath{{pics/}}

\begin{document}

\title{
Effect of transport current on suppression of superconductivity
with ultrashort laser pulse
}

\author{P F Kartsev, I O Kuznetsov}
\address{National Research Nuclear University MEPhI 
(Moscow Engineering Physics Institute), 
115409, Russian Federation, Moscow, Kashirskoe hwy, 31}


\begin{abstract}
We study the suppression of superconductivity
with ultrashort laser pulse in the presence of transport current.
The theoretical model is based on the BCS relations for the superconducting state
coupled with kinetic equations for nonequilibrium
Bogoliubov quasiparticles and phonons.
The results of numerical simulation for 
picosecond and femtosecond laser pulses
of optical and infrared ranges are given.
We discuss the effects of main problem parameters,
including the current density.
\end{abstract}


\maketitle

\section{Introduction}  

There exists the important task 
of developing fast switching devices for use 
in high-current superconducting applications,
e.g. current limiters \cite{Jin2012},
NMR/MRI tomography magnets  \cite{qu2016},
magnetic levitation devices \cite{terai2006},
inductive pulsed power supplies \cite{wu2013} and others.
There are several control principles
used in superconducting switches:
thermal \cite{kim2015}, current-based \cite{osipov2019},
magnetic \cite{sung2009,jaeyoungjang2010},
laser \cite{gray1984}
or combinations \cite{anischenko2019}.
As the creation of current or magnetic field pulses
shorter than $\sim$10$^{-6}$ seconds is difficult,
using the laser for fast switching seems the most preferable.
The estimation of the shortest switching time
is given by the relaxation time of electron subsystem in superconductors
$\sim$10$^{-13}$ seconds and lower,
and can be possibly achieved with femtosecond pulses \cite{kabanov2005}.

The nonequilibrium state of excess quasiparticles and phonons 
in the superconductor can be treated differently,
depending on the duration of the excitation.
In the case of source with slowly changing or constant intensity,
the stationary nonequilibrium distribution function
is usually considered,
depending on the excitation parameters such as power, spectral width etc.
 \cite{Elesin, Kashurnikov_C, Kashurnikov_PT}.
 One of the prominent results 
 is the superconductivity stimulation by the
 microwave field \cite{eliashberg}.
On the other hand, to study the strongly nonequilibrium state
after nearly-instant excitation
e.g. absorption of gamma radiation quantum \cite{Ovchinnikov},
 the more general approaches are preferable,
 such as
the time-dependent Green's functions method \cite{timeDependentGreenFunctions},
Keldysh formalism \cite{Keldysh, LL_PhysKin},
and Eliashberg's theory \cite{eliashberg_review}.

The excitation of superconductor 
with ultrashort laser pulse discussed in this work,
corresponds to the duration intermediate between these extreme cases.
Following the typical experimental results \cite{kusar, long_2006},
there are two distinct stages in the reaction of the superconductor.
First, the deviation from the equilibrium
 develops further after the excitation 
in the sub-picosecond time scale, 
which is governed by the electron-electron interaction.
After that, the superconducting state
 returns relatively slowly to the equilibrium
with characteristic times of tens of picoseconds,
due to the electron-phonon interaction.
The relaxation 
of the excess nonequilibrium quasiparticles and phonons
can be described by the phenomenological
Rothwarf-Taylor model \cite{RT, kabanov2005}. 

In this work, the duration of superconductivity suppression process
determining the potentially achievable operating speed, is studied.
We use the qualitative theoretical model
based on the 
Bardeen-Cooper-Shrieffer (BCS) theory 
coupled with kinetic equations for high-energy 
quasiparticles and phonons,
demonstrating the experimentally known characteristic behaviour
with fast stage of the excitation development.
The results of the numerical simulation 
for typical problem parameters are presented.
The dependence of the superconductivity suppression time
on the current density and laser wavelength is discussed.

\section{Theoretical model}  

We consider the limited area of superconductor
irradiated by the optical pulse of finite duration
generating the nonequilibrium Bogoliubov quasiparticles.
The recombination of two quasiparticles
into a Cooper pair creates the phonon carrying the extra energy,
while the backwards process converts the phonon and Cooper pair 
into two quasiparticles.
This energy exchange between subsystems
allows the thermalization
of the excited nonequilibrium particle densities 
at relatively short times.
After that, the excess quasiparticles and phonons 
decay with corresponding characteristic times
resulting in the relaxation of the superconducting state.

To affect the superconducting order parameter,
the quasiparticle energy should be lower than Debye energy
($|\xi| < \hbar \omega_{\mathrm{D}} \sim 10^{-2}$~eV).
However, the quasiparticles generated 
by the absorption of the optical quantum of frequency $\omega$
have much larger energy $\simeq\hbar \omega/2 \sim$~1eV
and thus need some time to reach the effective range with relaxation.
We should mention the special case of
THz radiation \cite{cavalleri2016},
in which the radiation quantum energy is comparable to the energy gap
and the generated quasiparticles affect the superconductivity immediately.
In this work, we limit the consideration to the 
laser pulses of optical and near-infrared ranges.

Such formulation allows to avoid the difficulties
of applying the theoretical approach based on Green's functions
 \cite{Ovchinnikov,gulyan1981}.
Indeed, as the superconducting state 
is affected by the quasiparticles only near the Fermi surface,
the optically-generated excitations have much higher energy
and can be described with simple kinetic equations
coupled with BCS relations for the order parameter 
in a self-consistent way.
This approach is feasible
at time scales long enough for the BCS state 
to adjust to the varying number of excitations,
i.e.   $\gtrsim$10$^{-14}$~s.

Main relations of the BCS theory:
 \begin{eqnarray}
 \label{delta_sc}
 \Delta = V_0 {\sum \limits_{{\bf k}} }^{\prime}
 u_{\bf k} v_{\bf k} (1 - n_{{\bf k},\uparrow} - n_{-{\bf k},\downarrow} ),
 \\
 \label{u_sc}
 u_{\bf k}^2= \frac{1}{2} \left( 1 + \frac{\xi_{\bf k}}{\varepsilon_{\bf k}}\right),
 \\
 \label{v_sc}
 v_{\bf k}^2= \frac{1}{2} \left( 1 - \frac{\xi_{\bf k}}{\varepsilon_{\bf k}}\right),
 \\
 \label{xi_sc}
 \xi_{\bf k} = \frac{\hbar^2 {\bf k}^2}{2 m} - E_{\mathrm{F}},
 \\
 \label{energy_sc}
 \varepsilon_{\bf k}= \sqrt{\xi_{\bf k}^2 + \Delta^2},
 \end{eqnarray}
where
$V_0$ is the interaction parameter,
$n_{{\bf k}\uparrow}$ and $n_{{\bf k}\downarrow}$
are the numbers of electrons with wave number ${\bf k}$
and the respective spin projections, 
the dash at the sum denotes the restriction for the electron energies
$|\xi_{\bf k}|<\hbar \omega_{\mathrm{D}}$,
$\omega_{\mathrm{D}}$ is the Debye frequency,
$m$ is the electron mass,
 $E_{\mathrm{F}}$ is the Fermi energy,
and $\Delta$ is the superconducting energy gap (order parameter).

The elementary excitations in a superconductor
are the Bogoliubov quasiparticles
with 
energies $\varepsilon_{\bf k}$ and
secondary quantization operators
$\hat \gamma^\dagger_{{\bf k}\sigma}$, $\hat \gamma_{{\bf k}\sigma}$
related to the initial electron operators
$\hat a^\dagger_{{\bf k}\sigma}$, $\hat a_{{\bf k}\sigma}$
as follows:
 \begin{eqnarray}
\hat a_{{\bf k}\uparrow} = 
u^{*}_{\bf k} \hat \gamma_{{\bf k}\uparrow}
+ 
v_{\bf k} \hat \gamma^\dagger_{-{\bf k}\downarrow},
\\
\nonumber
\hat a^\dagger_{-{\bf k}\downarrow} = 
- 
v^{*}_{\bf k} \hat \gamma_{{\bf k}\uparrow}
+
u_{\bf k} \hat \gamma^\dagger_{-{\bf k}\downarrow}.
 \end{eqnarray}

Writing the Hamiltonian of the system in the quasiparticle basis 
and adding the electron-electron and electron-phonon interactions:
\begin{eqnarray}
\label{hamiltonian}
\hat H = 
 \sum \limits_{{\bf k}\sigma} \varepsilon_{\bf k} 
  n_{{\bf k}\sigma}
+
 \sum \limits_{\bf q} \tilde \varepsilon_{\bf q} 
  {\tilde n}_{\bf q}
+
\hat H_{\mathrm{e-e}}
+ 
\hat H_{\mathrm{e-ph}},
\\
\nonumber
\hat H_{\mathrm{e-e}} = 
 U_0 
\sum \limits_{ {\bf kmp} }
	\hat \gamma^\dagger_{{\bf k}\uparrow} \hat \gamma^\dagger_{{\bf m}\downarrow}
	\hat \gamma_{{{\bf m}+{\bf p}}\downarrow} \hat \gamma_{{\bf k}-{\bf p}\uparrow},
\\
\nonumber
\hat H_{\mathrm{e-ph}} =
 M_0 
\sum \limits_{{\bf kq}\sigma} 
	(
 \hat b^\dagger_{\bf q}
		\hat a^\dagger_{{\bf k}\sigma} \hat a_{{\bf k}+{\bf q},\sigma}
	 + H.c.),
\end{eqnarray}
where  
$\hat b^\dagger_{\bf q}$, $\hat b_{\bf q}$
denote the phonon creation and annihilation operators,
$n_{{\bf k}\sigma} =
\hat \gamma^\dagger_{{\bf k}\sigma} \hat \gamma_{{\bf k}\sigma}$,
${\tilde n}_{\bf q} =
\hat b^\dagger_{\bf q} \hat b_{\bf q}$
are the occupation numbers of quasiparticles and phonons, correspondingly,
and $\varepsilon_{\bf k}$ and $\tilde \varepsilon_{\bf q}$
are the dispersion relations of quasiparticles and phonons.
We use
the matrix elements of electron-electron and electron-phonon interactions
$U_0$ and $M_0$ independent of momentum,
as the specific form of the interaction terms is not critical
for the behaviour studied in this work.
We also assume the laser radiation uniform,
to make the work values coordinate-independent.

Using the gauge in which the parameters
$u_{\bf k}$, $v_{\bf k}$ are real-valued,
and the fact $u_{\bf k}=u_{-{\bf k}}$, $v_{\bf k}=v_{-{\bf k}}$,
we convert the last term in  (\ref{hamiltonian}) 
to the quasiparticle basis:
\begin{eqnarray}
\hat H_{\mathrm{e-ph}}
= M_0 
\sum \limits_{\bf kq}
	\left[
		\hat b^\dagger_{\bf q}
		\hat \gamma^\dagger_{{\bf k}\uparrow} \hat \gamma_{{\bf k}+{\bf q},\uparrow}
		(u_{\bf k} u_{{\bf k}+{\bf q}} - v_{\bf k} v_{{\bf k}+{\bf q}})
		\right.
\\
\nonumber
		+
		\hat b^\dagger_{\bf q}
		\hat \gamma^\dagger_{{\bf k}\downarrow} \hat \gamma_{{\bf k}+{\bf q},\downarrow}
		(u_{\bf k} u_{{\bf k}+{\bf q}} - v_{\bf k} v_{{\bf k}+{\bf q}})
\\
\nonumber
		+
		\hat b^\dagger_{\bf q}
		\hat \gamma_{{\bf k},\downarrow} \hat \gamma_{{\bf q}-{\bf k},\uparrow}
		(u_{\bf k} v_{{\bf q} - {\bf k}} + v_{\bf k} u_{{\bf q}-{\bf k}})
\\
\nonumber
		+
		\left.
		\hat b^\dagger_{\bf q}
		\hat \gamma^\dagger_{{\bf k},\uparrow} \hat \gamma^\dagger_{-{\bf k}-{\bf q},\downarrow}
		(u_{\bf k} v_{{\bf k} + {\bf q}} + v_{\bf k} u_{{\bf k}+{\bf q}})
	 \right] + H.c.
\end{eqnarray}
The first two lines correspond to the 
scattering of the quasiparticle creating the phonon,
the third line
describes the recombination of two quasiparticles creating the phonon,
while the fourth gives the obviously unphysical process
of three-particle creation without energy conservation.
It is useful to introduce the coefficients
for the scattering and recombination processes:
\begin{eqnarray}
S_{\bf kq} = M_0
		\left( u_{\bf k} u_{{\bf k}+{\bf q}} - v_{\bf k} v_{{\bf k}+{\bf q}} \right),
\\
R_{\bf kq} = M_0
		(u_{\bf k} v_{{\bf q} - {\bf k}} + v_{\bf k} u_{{\bf q}-{\bf k}}).
\end{eqnarray}

The system of kinetic equations \cite{kabanov2008} writes as:
\begin{eqnarray}
\label{dnq_dt}
\frac{ dn_{{\bf k}\sigma} }{dt}= 
- \frac{n_{{\bf k}\sigma} }{\tau_{\mathrm{e}} }
+ G_{{\bf k}\sigma} 
- R_{{\bf k}\sigma} 
+ J_{{\bf k}\sigma}^{\mathrm{(e-e)}} 
+ J_{{\bf k}\sigma}^{\mathrm{(S)}} 
+ J_{{\bf k}\sigma}^{\mathrm{(R)}},
\\
\label{dnph_dt}
\frac{d \tilde n_{\bf q} }{dt}= 
- \frac{ {\tilde n}_{\bf q} }{\tau_{\mathrm{ph}}}
+ {\tilde J}_{\bf k}^{\mathrm{(S)}} 
+ {\tilde J}_{\bf k}^{\mathrm{(R)}}.
\end{eqnarray}

The lifetimes $\tau_{\mathrm{e}}$, $\tau_{\mathrm{ph}}$ 
of nonequilibrium quasiparticles and phonons
correspond to leaving the excited area.

The optical generation and recombination rates $G$, $R$
describing the creation or annihilation 
of two quasiparticles with momenta ${\bf k}$, ${- \bf k}$
are proportional to the radiation intensity 
at the corresponding frequency  $\omega$
and depend on the particle occupations:
\begin{eqnarray}
\label{temp_G}
G_{{\bf k}\sigma}  = 
\alpha
I \left( \frac{2 \varepsilon_{\bf k}}{\hbar} \right) 
(1-n_{{\bf k}\sigma}) (1-n_{-{\bf k},-\sigma}),
\\
\label{temp_R}
R_{{\bf k}\sigma}  = 
\alpha
I \left( \frac{2 \varepsilon_{\bf k}}{\hbar} \right) 
n_{{\bf k}\sigma} n_{-{\bf k},-\sigma},
\end{eqnarray}
where $I(\omega)$ is the intensity
and $\alpha$ is the prefactor.

The collision integrals for various interaction terms
have the form \cite{ICMSquare_2017}:
\begin{eqnarray}
\nonumber
J_{{\bf k}\sigma}^{\mathrm{(e-e)}} = 
\frac{2\pi}{\hbar}
U_0^2
\sum \limits_{{\bf p m r}}
\left[
(1-n_{{\bf k}\sigma}) (1-n_{{\bf p},-\sigma}) 
   n_{{\bf m},-\sigma} n_{{\bf r}\sigma}
\right.
\\
\label{J_ee}
- \left.
n_{{\bf k}\sigma} n_{{\bf p},-\sigma} 
 (1-n_{{\bf m},-\sigma}) (1-n_{{\bf r}\sigma}) 
\right]
 F(\Delta \varepsilon) \delta_{{\bf k}+{\bf p},{\bf m}+{\bf r}},
 \\
 \nonumber
J_{{\bf k}\sigma}^{\mathrm{(S)}} = 
\frac{2\pi}{\hbar}
\sum \limits_{\bf q}
|S_{\bf kq}|^2
\left[
(1-n_{{\bf k}\sigma})n_{{\bf k}-{\bf q},\sigma} \tilde n_{\bf q} 
F(\varepsilon_{\bf k} - \varepsilon_{{\bf k}-{\bf q}} - \tilde \varepsilon_{\bf q} )
\right.
\\
\nonumber
+ (1-n_{{\bf k}\sigma})n_{{\bf k}+{\bf q},\sigma}
 (\tilde n_{\bf q} + 1 )
F(\varepsilon_{\bf k} 
- \varepsilon_{{\bf k}+{\bf q}} + \tilde \varepsilon_{\bf q} )
\\
\nonumber
- n_{{\bf k}\sigma} (1-n_{{\bf k}-{\bf q},\sigma}) (\tilde n_{\bf q} + 1) 
F(\varepsilon_{\bf k} - \varepsilon_{{\bf k}-{\bf q}} - \tilde \varepsilon_{\bf q} )
\\
\label{J_e_S}
-
\left.
n_{{\bf k}\sigma} (1-n_{{\bf k}+{\bf q},\sigma}) \tilde n_{\bf q}
F(\varepsilon_{\bf k} - \varepsilon_{{\bf k}+{\bf q}} + \tilde \varepsilon_{\bf q} )
\right],
\\
\nonumber
J_{{\bf k}\sigma}^{\mathrm{(R)}} = 
\frac{2\pi}{\hbar}
\sum \limits_{\bf q}
|R_{\bf kq}|^2
[
(1-n_{{\bf k}\sigma})(1-n_{{\bf q}-{\bf k},-\sigma}) \tilde n_{\bf q} 
\\
-
 n_{{\bf k}\sigma} n_{{\bf q}-{\bf k},-\sigma}
 (\tilde n_{\bf q} + 1 )
]
F(\varepsilon_{\bf k} + \varepsilon_{{\bf q}-{\bf k}} - \tilde \varepsilon_{\bf q} )
\label{J_e_R}
\end{eqnarray}
for quasiparticles,
\begin{eqnarray}
\nonumber
{\tilde J}_{\bf k}^{\mathrm{(S)}} = 
\frac{2\pi}{\hbar}
\sum \limits_{\bf k \sigma}
|S_{\bf kq}|^2
\left[
(1-n_{{\bf k}\sigma})n_{{\bf k}+{\bf q},\sigma}
 (\tilde n_{\bf q} + 1 )
F(\varepsilon_{\bf k} - \varepsilon_{{\bf k}+{\bf q}} + \tilde \varepsilon_{\bf q} )
\right.
\\
- \left.
(1-n_{{\bf k}\sigma})n_{{\bf k}-{\bf q},\sigma} \tilde n_{\bf q} 
F(\varepsilon_{\bf k} - \varepsilon_{{\bf k}-{\bf q}} - \tilde \varepsilon_{\bf q} )
\right],
\label{J_ph_S}
\\
\nonumber
{\tilde J}_{\bf k}^{\mathrm{(R)}} = 
\frac{2\pi}{\hbar}
\sum \limits_{\bf k}
|R_{\bf kq}|^2
F(\varepsilon_{\bf k} + \varepsilon_{{\bf q}-{\bf k}} - \tilde \varepsilon_{\bf q} )
\\
\times
[
 n_{{\bf k}\sigma} n_{{\bf q}-{\bf k},-\sigma}
 (\tilde n_{\bf q} + 1 )
 -
(1-n_{{\bf k}\sigma})(1-n_{{\bf q}-{\bf k},-\sigma}) \tilde n_{\bf q} 
]
\label{J_ph_R}
\end{eqnarray}
and phonons, correspondingly,
and the factor  $F( \Delta \varepsilon )$ was introduced to
account the broadening of energy levels.

\section{Numerical simulation}  

In the simulation, the following parameter values were used:
Debye temperature  T$_{\mathrm D}$$\simeq$350~K
(k$_{\mathrm B}$T$_{\mathrm D}$=$\hbar \omega_{\mathrm D}$=0.03~eV), 
Fermi energy E$_{\mathrm{ F}}$=3~eV, 
Fermi wave number k$_{\mathrm F}$=0.75$k_{\mathrm{max}}$,
where $k_{\mathrm{max}}=\pi/a$ is the boundary of the 1st Brillouin zone.
Initial value of the energy gap
$\Delta$(t=0)=10$^{-3}$~eV$\sim$k$_{\mathrm{B}}$T$_c$.
The values belong to the range typical to superconductors:
e.g. Nb
(T$_{\mathrm D}$=275~K, E$_{\mathrm{ F}}$=6.5~eV, T$_c$=9.2~K),
Nb$_3$Sn \cite{Nb3Sn_parameters}
(T$_{\mathrm D}$=300~K, E$_{\mathrm{ F}}$=10.5~eV, T$_c$=18.9~K),
La-Sr-Cu-O \cite{LSCO_parameters}
(T$_{\mathrm D}$=310$\dots$370~K, $E_{\mathrm{F}}$=0.1~eV, $T_c\simeq$40~K).
Worthy to note that
the specific numbers are not essential for the phenomena under study
as far as the main relation 
$\Delta \ll  \hbar \omega_{\mathrm D} \ll E_{\mathrm{ F}}$
is satisfied.

The value of matrix element $V_0$
is calculated following the BCS equation (\ref{delta_sc}).
The matrix element of electron-phonon interaction $M_0$
is derived from the time between electron-lattice collisions
$\tau \sim \hbar /2\pi M_0^2$ which 
for the typical metal at the temperature $T \sim 1$~K
can be esimated $\sim$10$^{-11}$~s.
The matrix element of electron-electron interaction $U_0$
was taken equal to $M_0$.
As the ratio between electron-electron and electron-phonon
  interaction constants 
for different superconducting materials varies widely
\cite{Elesin,Kashurnikov_C},
the value $C \simeq 1$ can be considered realistic.

The time of electron-electron scattering, however,
not only is determined by the material parameters
but also depends nonlinearly on the carrier density.
As a result, it can be as large as picoseconds
 \cite{times_htsc_eph,times_table}
and as low as attoseconds \cite{times_attosecond_metals}
for different materials.
For the problem considered in this work,
the value of electron-electron scattering time is determined
by the density of quasiparticles 
corresponding to the absorbed laser energy.

The intensity of optical radiation $I$
is measured in units giving the coefficient
$\alpha=1$ in equations (\ref{temp_G}), (\ref{temp_R})
i.e. in number of quanta absorbed,
and its time dependence is approximated by Gaussian function
$\sim \exp(-t^2/2 \sigma^2 )$ with duration given by $\sigma$.
We assume that the laser pulse uniformly illuminates
the whole area of consideration
with linear size 10~$\mu$m.
Then, using the values of Fermi velocity $\sim$10$^6$~m/s
and sound velocity $\sim$10$^3$~m/s, 
we can estimate the lifetimes of quasiparticles and phonons
as the times of leaving the system
$\tau_{\mathrm{e}}=$10$^{-11}$~s and
 $\tau_{\mathrm{ph}}=$10$^{-8}$~s, correspondingly.

In our study, we consider the temperature low enough
for the equilibrium densities of
quasiparticles and phonons to be zero.
With the nonequilibrium quasiparticles present,
the value of order parameter $\Delta$ 
decreases following the equation (\ref{delta_sc}).
The simulation of the kinetic equations
(\ref{dnq_dt})---(\ref{J_ph_R})
is performed on the momentum lattice 32$\times$32$\times$32
which can be considered large enough 
to describe the macroscopic behaviour.
The discreteness of the energy spectrum in the system 
is compensated with broadening of the energy levels
$\Delta \varepsilon \sim 10^{-3}$~eV
and choosing wider spectral line of laser radiation
$\hbar \Delta \omega$=0.1~eV.
The laser line shape
and level broadening factor $F(\Delta \varepsilon)$
are taken in the form of Gaussian functions.
The numerical method used for the simulation of kinetic equations
is described in details in \cite{arxiv}.

\section{Superconductivity suppression with laser pulse}
\label{4_simulation}

\begin{figure}[h!]
\begin{minipage}{0.5\linewidth}\begin{center}
\includegraphics[width=1.0\linewidth]{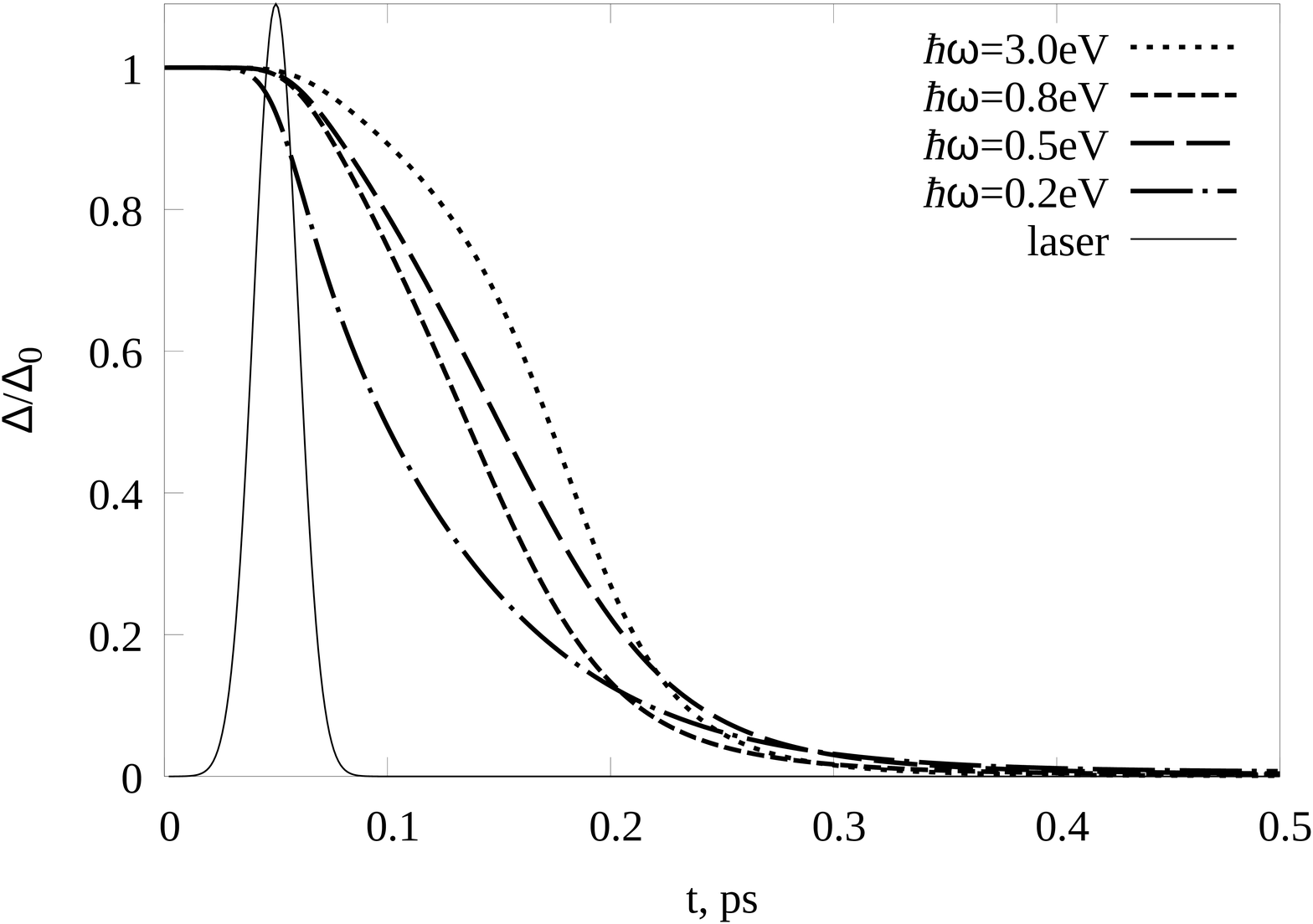}

(a)
\end{center}
\end{minipage}
\begin{minipage}{0.5\linewidth}\begin{center}

\includegraphics[width=1.0\linewidth]{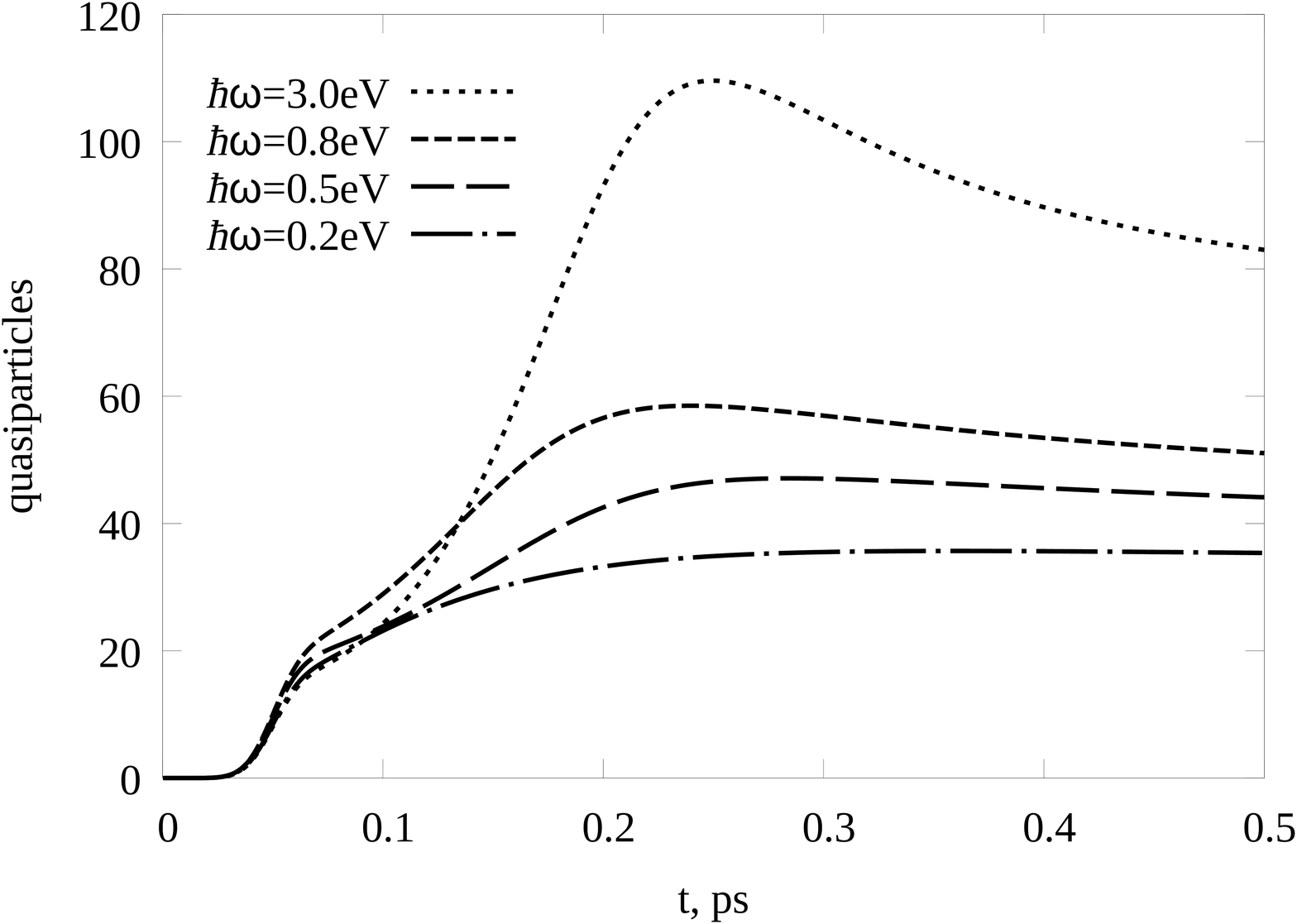}

(c)
\end{center}
\end{minipage}

\begin{minipage}{0.5\linewidth}\begin{center}

\includegraphics[width=1.0\linewidth]{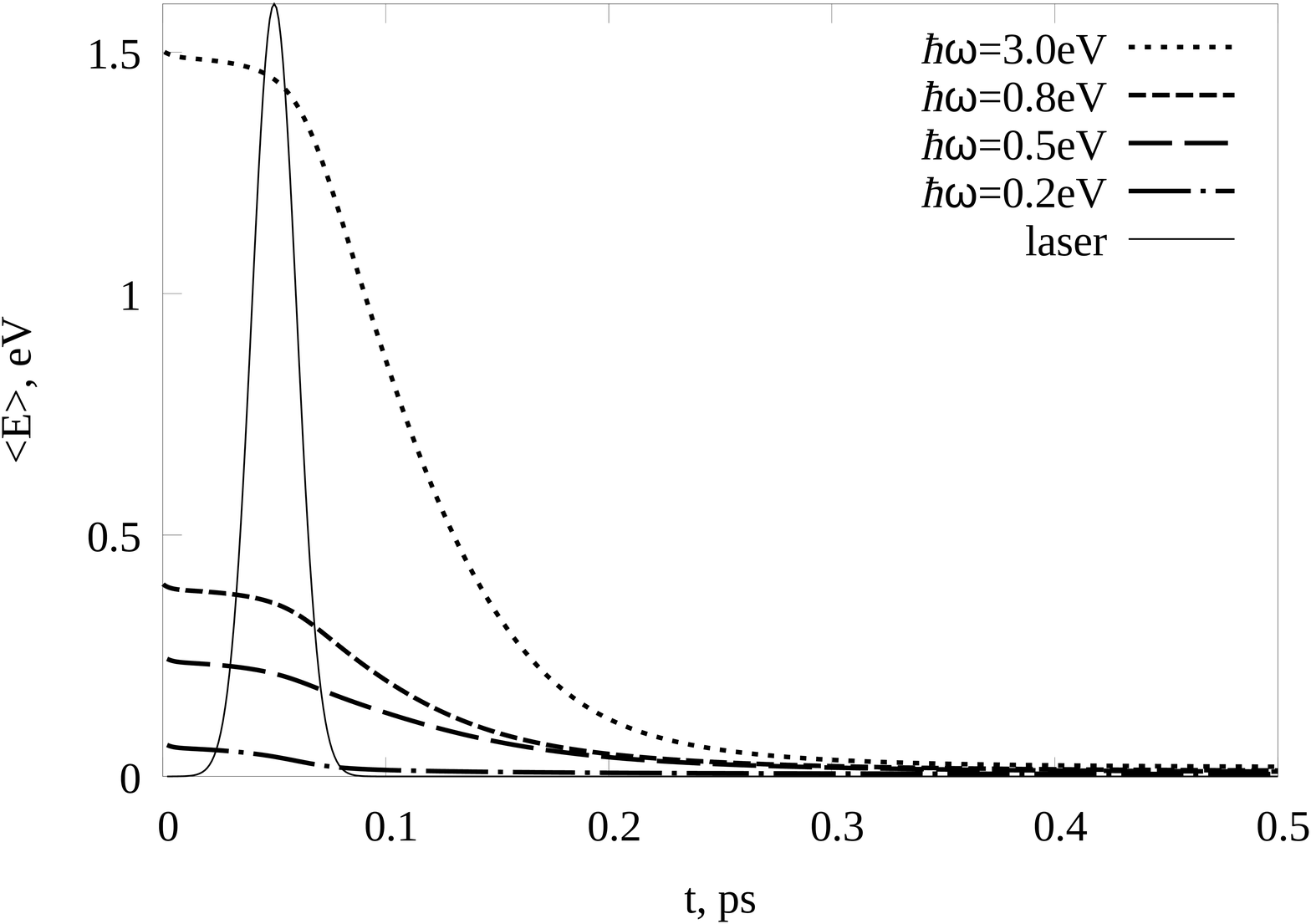}

(b)
\end{center}
\end{minipage}
\begin{minipage}{0.5\linewidth}\begin{center}

\includegraphics[width=1.0\linewidth]{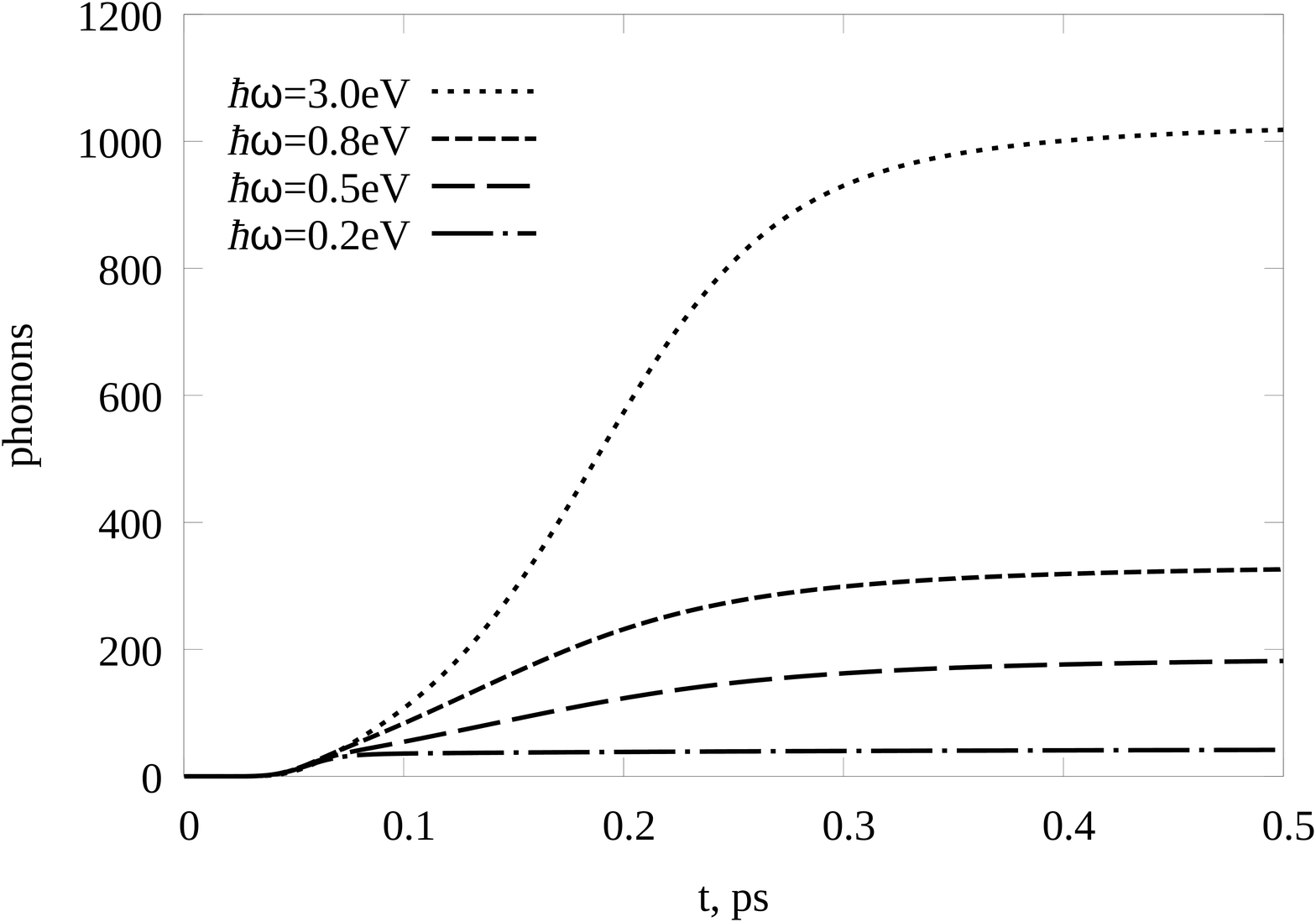}

(d)
\end{center}
\end{minipage}
\caption{
Results of simulation
for several energies of optical quantum 
$\hbar \omega$=0.2, 0.5, 0.8, 3.0~eV,
in the case of femtosecond pulse ($\sigma$=10~fs, $I$=1.0),
with (a) order parameter $\Delta$,
(b) average energy of quasiparticles,
(c) number of quasiparticles and (d) phonons.
}
\label{1_main_fs}
\end{figure}

\begin{figure}[h!]
\begin{minipage}{0.5\linewidth}\begin{center}
\includegraphics[width=1.0\linewidth]{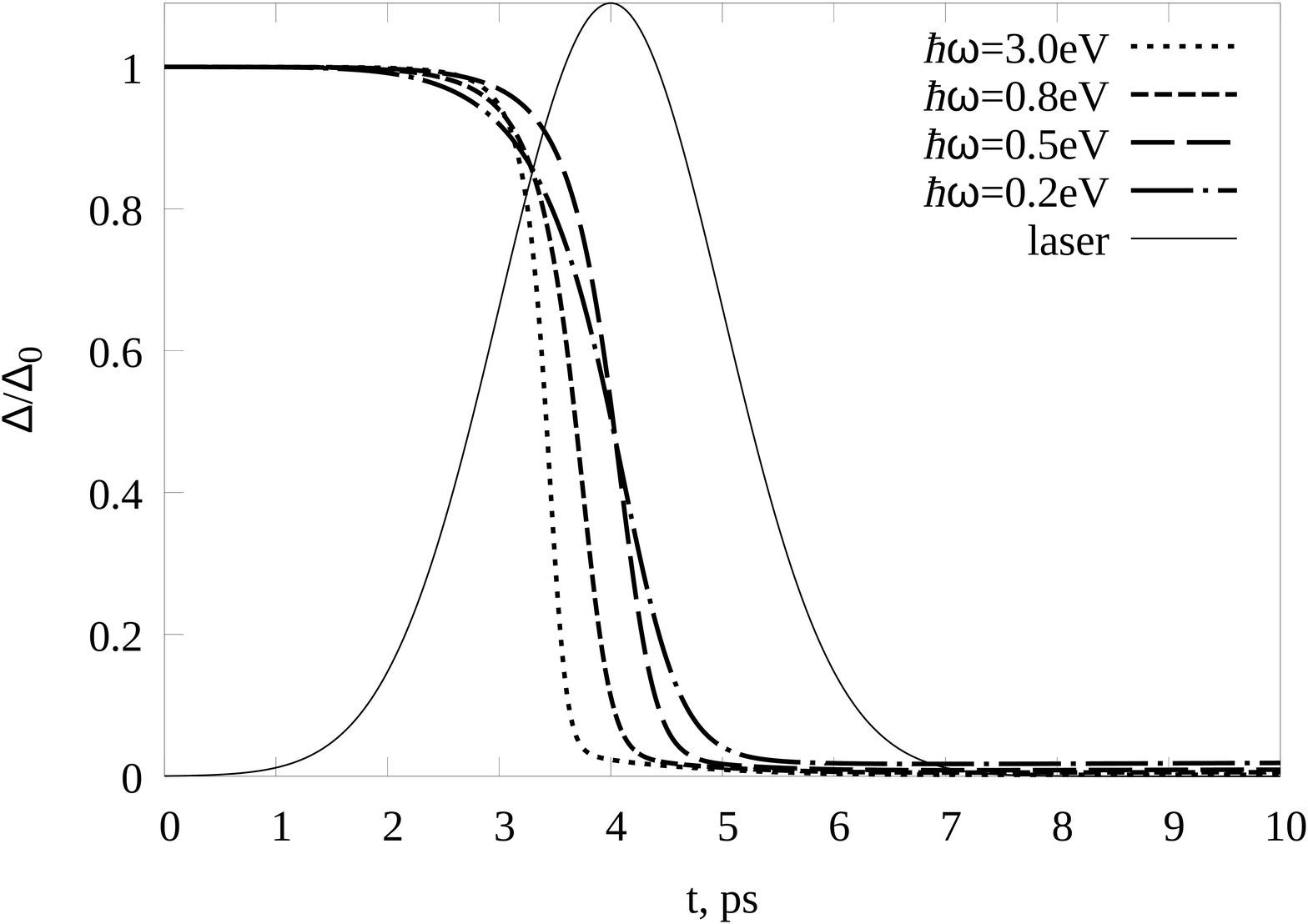}

(a)
\end{center}
\end{minipage}
\begin{minipage}{0.5\linewidth}\begin{center}

\includegraphics[width=1.0\linewidth]{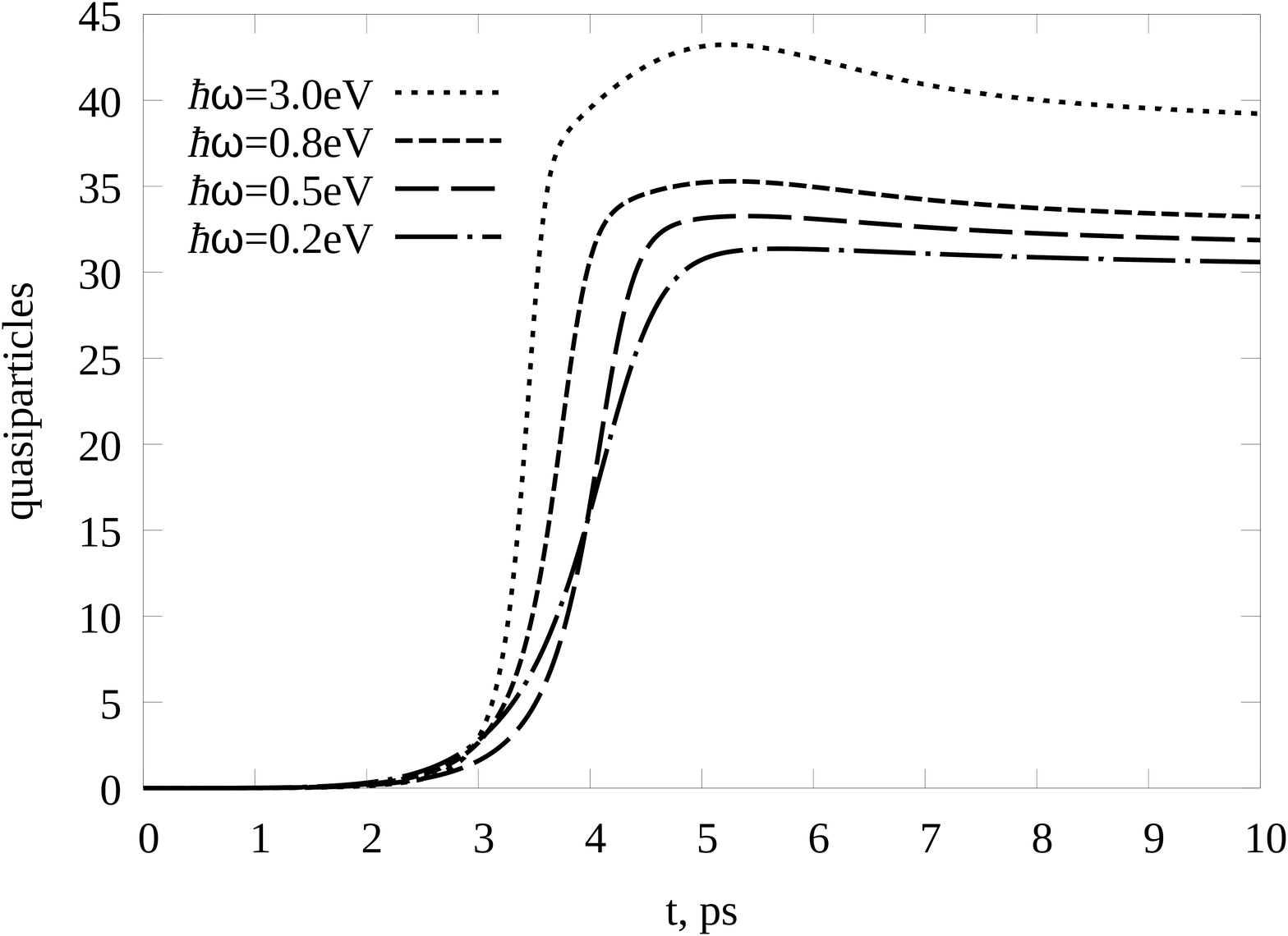}

(c)
\end{center}
\end{minipage}

\begin{minipage}{0.5\linewidth}\begin{center}

\includegraphics[width=1.0\linewidth]{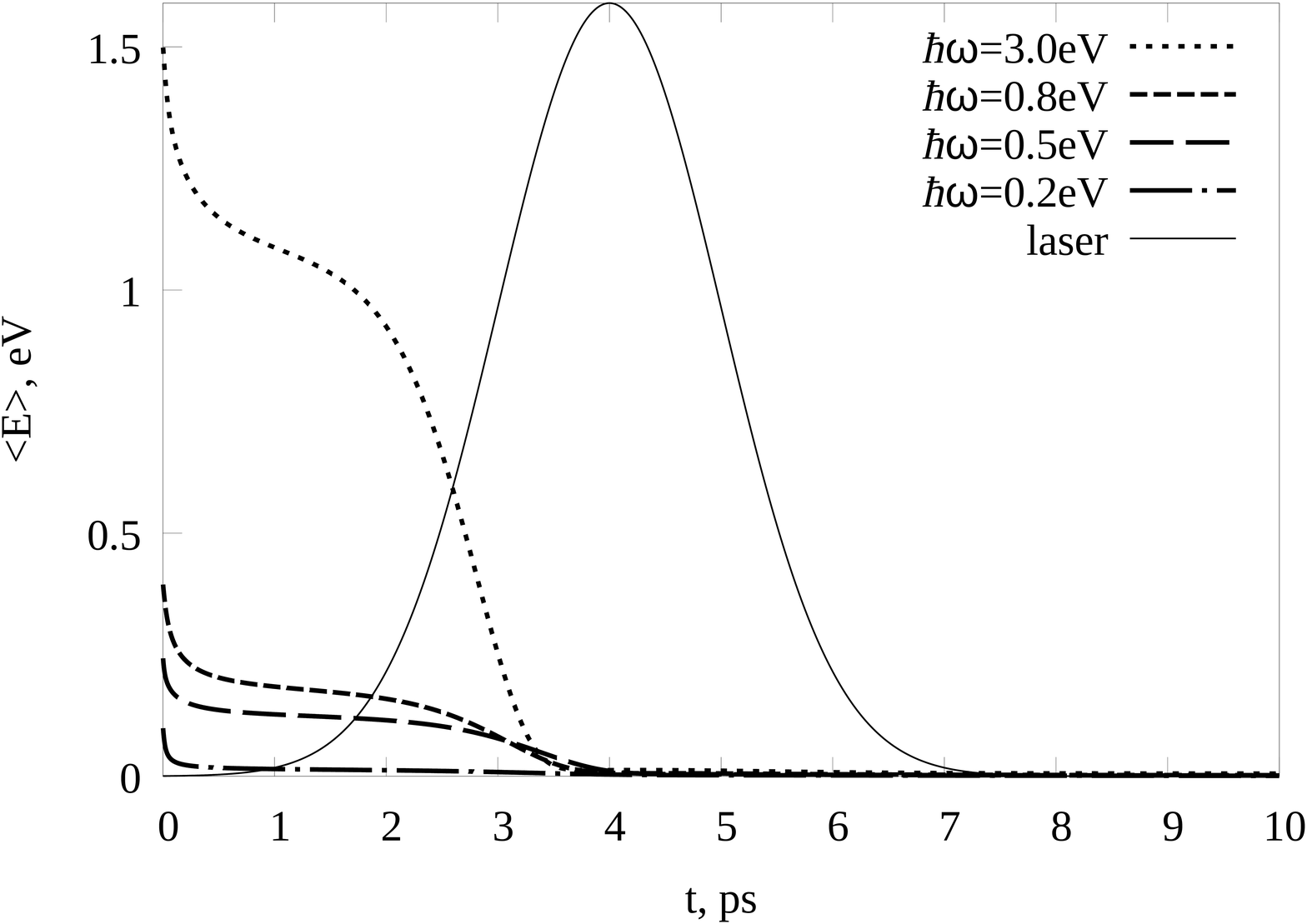}

(b)
\end{center}
\end{minipage}
\begin{minipage}{0.5\linewidth}\begin{center}

\includegraphics[width=1.0\linewidth]{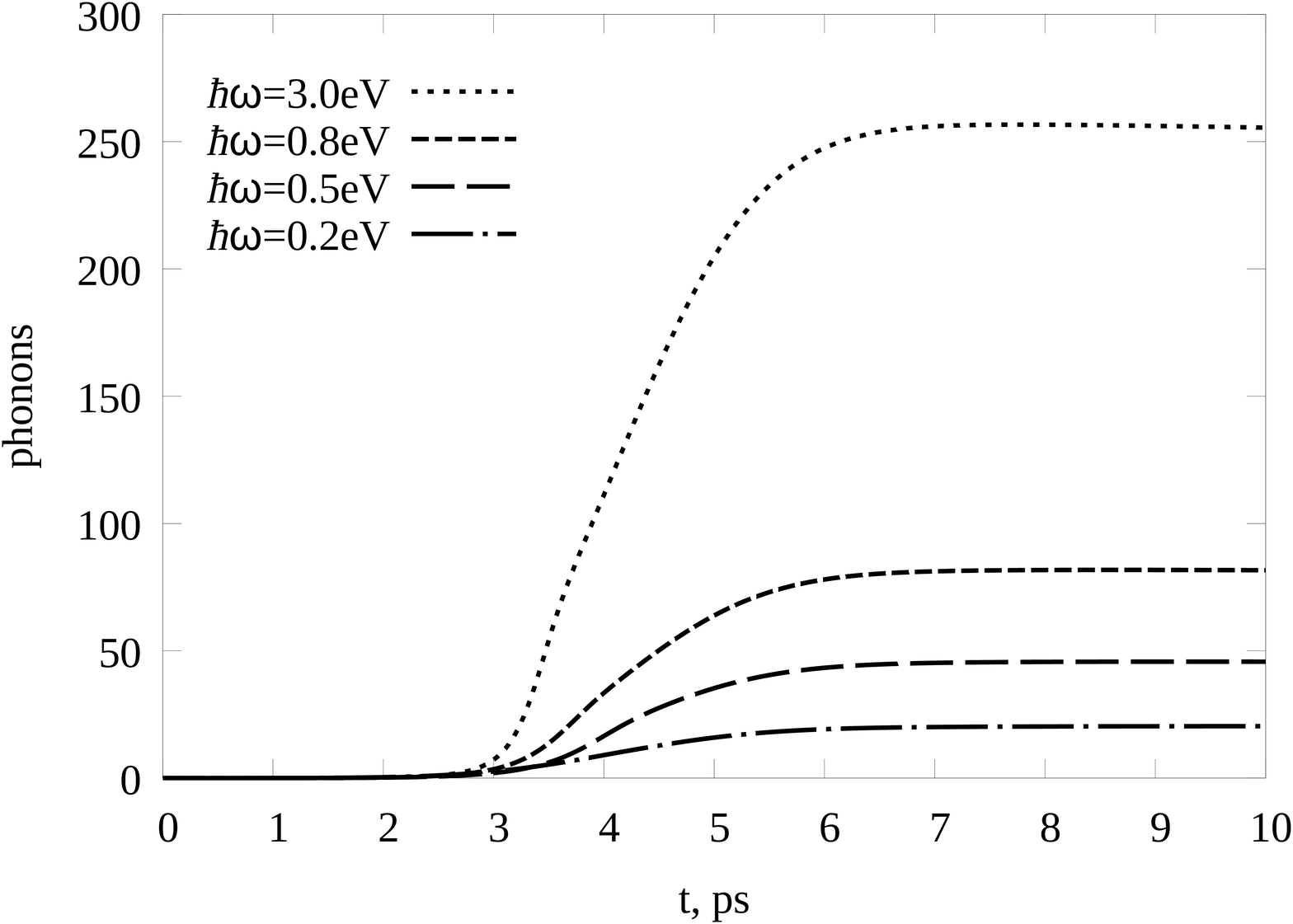}

(d)
\end{center}
\end{minipage}
\caption{
Same as figure \ref{1_main_fs}, for the picosecond pulse
($\sigma$=1~ps, $I$=2.5$\cdot$10$^{-3}$).
}
\label{1_main_ps}
\end{figure}

In figures \ref{1_main_fs} and \ref{1_main_ps},
we show the simulation results
demonstrating the system behaviour
under the action of laser pulse.

In the case of shorter femtosecond pulse (figure \ref{1_main_fs}),
the complete suppression of the superconductivity
is delayed in relation to the pulse action,
approximately at 0.3~ps for the chosen problem parameters.
It is caused by the finite time needed for 
the high-energy quasiparticles to arrive near the Fermi surface due to relaxation.
In the case of longer picosecond pulse (figure \ref{1_main_ps}),
the suppression takes approximately 1~ps
and the further energy absorption results only 
in the generation of additional phonons.

The energy of optical quantum $\hbar \omega$
affects the time of switching
and the maximal number of quasiparticles only marginally.
On the other hand, higher values of the radiation quantum energy
delay the start of the order parameter decreasing
 (figure \ref{1_main_fs}(a)).
It can be explained by the time required
for the quasiparticles to reach low enough energies.

\begin{figure}
\begin{center}
\includegraphics[width=0.5\linewidth]{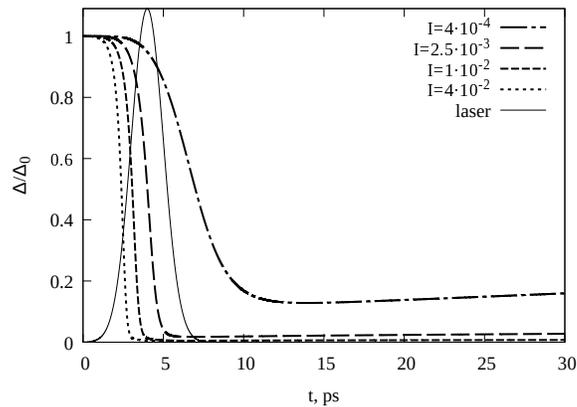}
\end{center}
\caption{
Suppression of the superconducting order parameter
with picosecond pulse ($\sigma$=1~ps, $\hbar \omega$=0.2~eV),
calculated for several intensities $I$.
}
\label{2_intensity}
\end{figure}

The effect of the higher intensity consists in the quicker 
suppression of superconductivity.
In figure  \ref{2_intensity}, we present
the results of calculation for the picosecond laser pulse
of various intensities.
We should note that for sufficiently large intensities, 
the profiles of the order parameter dropping are similar
due to nonlinear dependence of the relaxation rate on the particle density. 

Summarizing,
the superconductivity suppression in the discussed conditions
is related to several causes.
During the action of laser pulse,
the order parameter is depleted
due to energy absorption.
After that, the relaxation
of the excess high-energy quasiparticles
increases occupations
$n_{{\bf k}\uparrow}$, $n_{{\bf k}\downarrow}$
at energies lower that Debye energy
which enter the BCS equation (\ref{delta_sc})
and decrease the energy gap.

At large enough energy of the pulse,
the superconductivity can be switched off completely,
even during the time of laser pulse
(figure \ref{1_main_ps}).
This case, though, can lead to harmful side effects
due to the excess heat emission,
and slow down the backwards switching.
It makes preferable to choose the case of lower energy impact 
with delayed order parameter suppression.
Next we study the possibility to accelerate this stage of the process.

\section{Effect of transport current}  

The process of superconductivity suppression
after the action of ultrashort laser pulse 
can be affected by the transport current in two different ways.
First, the initial value of order parameter before the impact is lowered,
decreasing the energy needed to suppress the superconducting state
which can potentially provide the faster switching.
On the other hand, with the lower density of Cooper pairs
affecting the efficiency of the laser radiation absorption,
the decreased amount of quasiparticles is generated,
resulting in slower relaxation 
and delaying the suppression of the order parameter.
The relative contributions of these factors
depend on the radiation frequency
which determines the energy of the generated quasiparticles.

In the presence of trasport current,
the Cooper pairs have momentum $2\hbar {\bf k}_s$,
i.e. consist of electrons with momenta 
$\hbar({\bf k}_s+{\bf k})$ and $\hbar({\bf k}_s-{\bf k})$.
The unitary transformation coefficients
and the order parameter obtain the complex factor \cite{deGennes}:
\begin{eqnarray}
\nonumber
\Delta({\bf r}) = \Delta^{(0)} e^{2i {\bf k}_s {\bf r}},
\\
\nonumber
u_{\bf k}({\bf r}) = u^{(0)}_{\bf k}  e^{i {\bf k}_s {\bf r} },
\\
\nonumber
v_{\bf k}({\bf r}) = v^{(0)}_{\bf k}  e^{i {\bf k}_s {\bf r}}.
\end{eqnarray}
We can neglect the phase change 
over the linear size $L$ of the volume under consideration
if the current is small enough, so that
$k_s  L \ll 2\pi$, i.e.
$k_s \ll 2\pi/L \ll k_{\mathrm{F}}$.

In this case the relations 
 (\ref{u_sc})---(\ref{energy_sc})
remain unchanged
with equations
(\ref{delta_sc}), (\ref{xi_sc}), (\ref{temp_G}), (\ref{temp_R}) 
modified to use the corresponding momenta and particle numbers:
\begin{eqnarray}
\label{delta_current}
 \Delta = U_0 {\sum \limits_{{\bf k}} }^{\prime}
 u_{\bf k} v_{\bf k}
 	 (1 - n_{{\bf k}_s + {\bf k},\uparrow} - n_{{\bf k}_s - {\bf k},\downarrow} ),
\\
\label{xi_current}
\xi_{\bf k} = \frac{1}{2}\left( 
	\frac{\hbar^2 ({\bf k}_s+{\bf k})^2}{2m} + 
	\frac{\hbar^2 ({\bf k}_s-{\bf k})^2}{2m}
\right) - E_{\mathrm{F}},
\\
\label{temp_G_current}
G_{{\bf k}_s+{\bf k},\sigma}  = 
\alpha
I \left( \frac{\varepsilon_{{\bf k}_s+{\bf k}}+\varepsilon_{{\bf k}_s-{\bf k}}}{\hbar} \right) 
 (1-n_{{\bf k}_s+{\bf k},\sigma}) (1-n_{{\bf k}_s-{\bf k},-\sigma}),
\\
\label{temp_R_current}
R_{{\bf k}_s+{\bf k},\sigma}  = 
\alpha
I \left( \frac{\varepsilon_{{\bf k}_s+{\bf k}}+\varepsilon_{{\bf k}_s-{\bf k}}}{\hbar} \right) 
n_{{\bf k}_s+{\bf k},\sigma} n_{{\bf k}_s-{\bf k},-\sigma}.
\end{eqnarray}

The current in the superconductor 
is the sum of contributions from all electrons:
\begin{equation}                      
\nonumber
{\bf J} = \frac{e \hbar }{ m }
<\mathrm{BCS}|
\sum 
	\limits_{{\bf k} \sigma}
		{\bf k} \hat a^\dagger_{{\bf k}\sigma} \hat a_{{\bf k}\sigma}
|\mathrm{BCS}>
\end{equation}

This expression can be re-written as \cite{Abrikosov}:
\begin{equation}
\label{current}
 {\bf J} = 
\frac{e \hbar }{ m }
\sum 
	\limits_{{\bf k}}
		({\bf k}_s+{\bf k}) 
\left[
			|u_{\bf k}|^2
				   n_{ {\bf k}_s + {\bf k}}
			+ |v_{\bf k}|^2
				(2-n_{ {\bf k}_s - {\bf k}})
\right],
\end{equation}
where 
$n_{{\bf k}} = <\mathrm{BCS}| 
\sum \limits_{\sigma}
	\hat \gamma^\dagger_{ {\bf k}\sigma} \hat \gamma_{ {\bf k}\sigma}
|\mathrm{BCS}>$ is the number of quasiparticles with momentum
${\bf k}$
and the summation 
in (\ref{current})
goes over all momentum states in the Brillouin zone.

During the simulation,
the kinetic equations 
(\ref{dnq_dt})---(\ref{J_ph_R}) are solved numerically
along with equations
(\ref{delta_current})---(\ref{current})
to update the superconducting order parameter and
the momentum of Cooper pairs,
taking into account the occupations of low-energy quasiparticle levels.

\begin{figure}
\begin{minipage}{0.5\linewidth}\begin{center}
\includegraphics[width=1.0\linewidth]{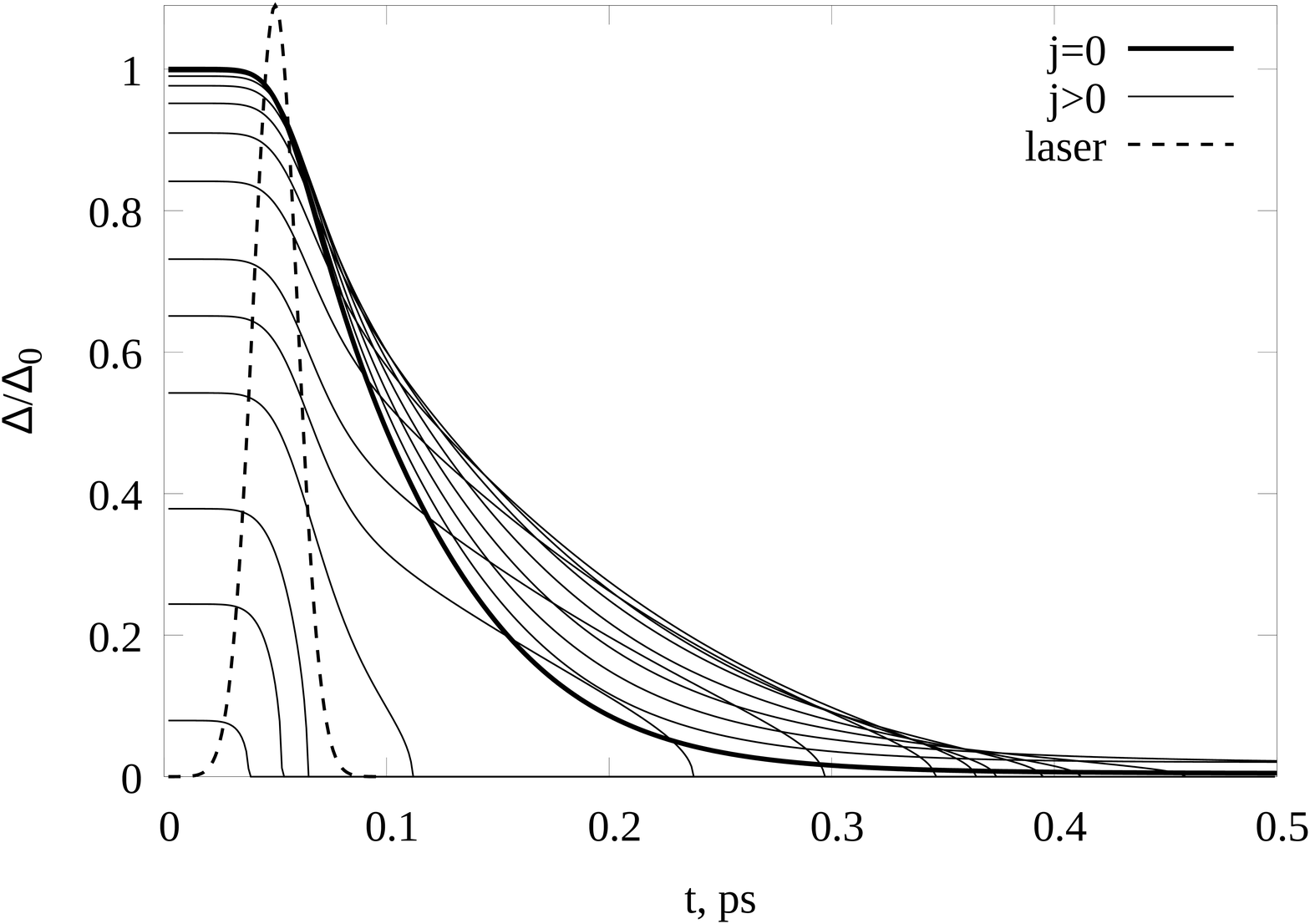}

(a) 
\end{center}
\end{minipage}
\begin{minipage}{0.5\linewidth}\begin{center}
\includegraphics[width=1.0\linewidth]{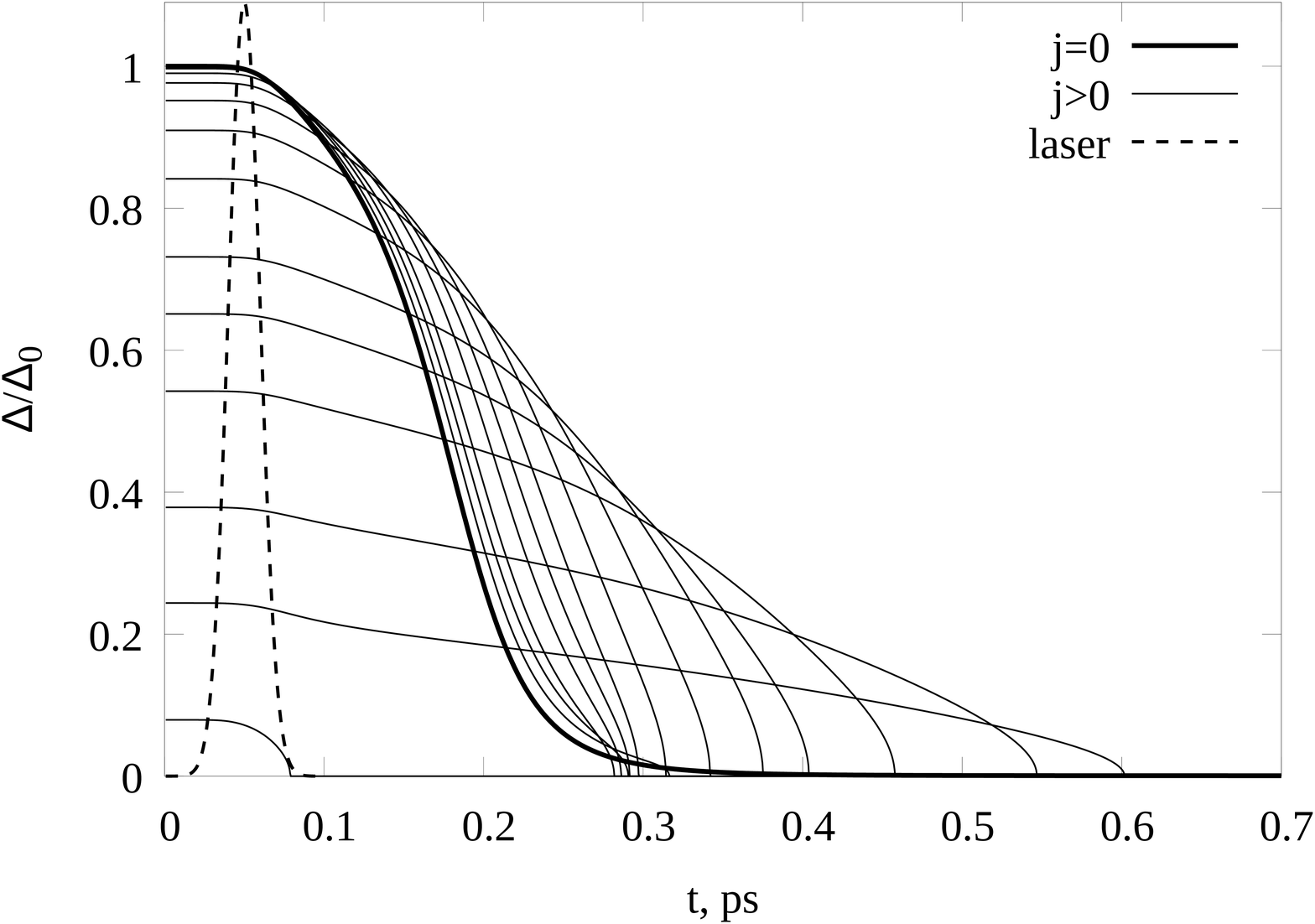}

(d) 
\end{center}
\end{minipage}

\begin{minipage}{0.5\linewidth}\begin{center}
\includegraphics[width=1.0\linewidth]{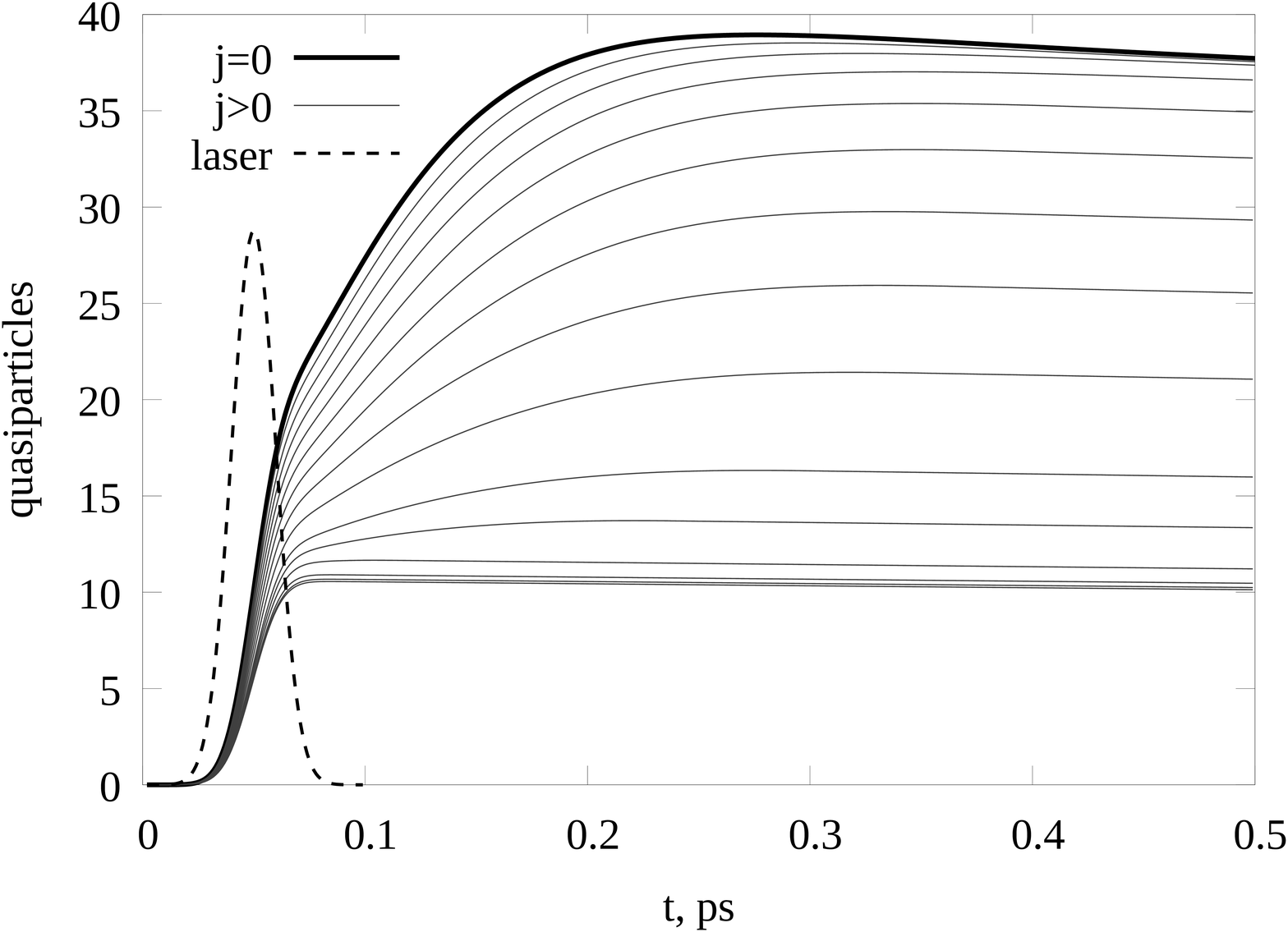}

(b) 
\end{center}
\end{minipage}
\begin{minipage}{0.5\linewidth}\begin{center}
\includegraphics[width=1.0\linewidth]{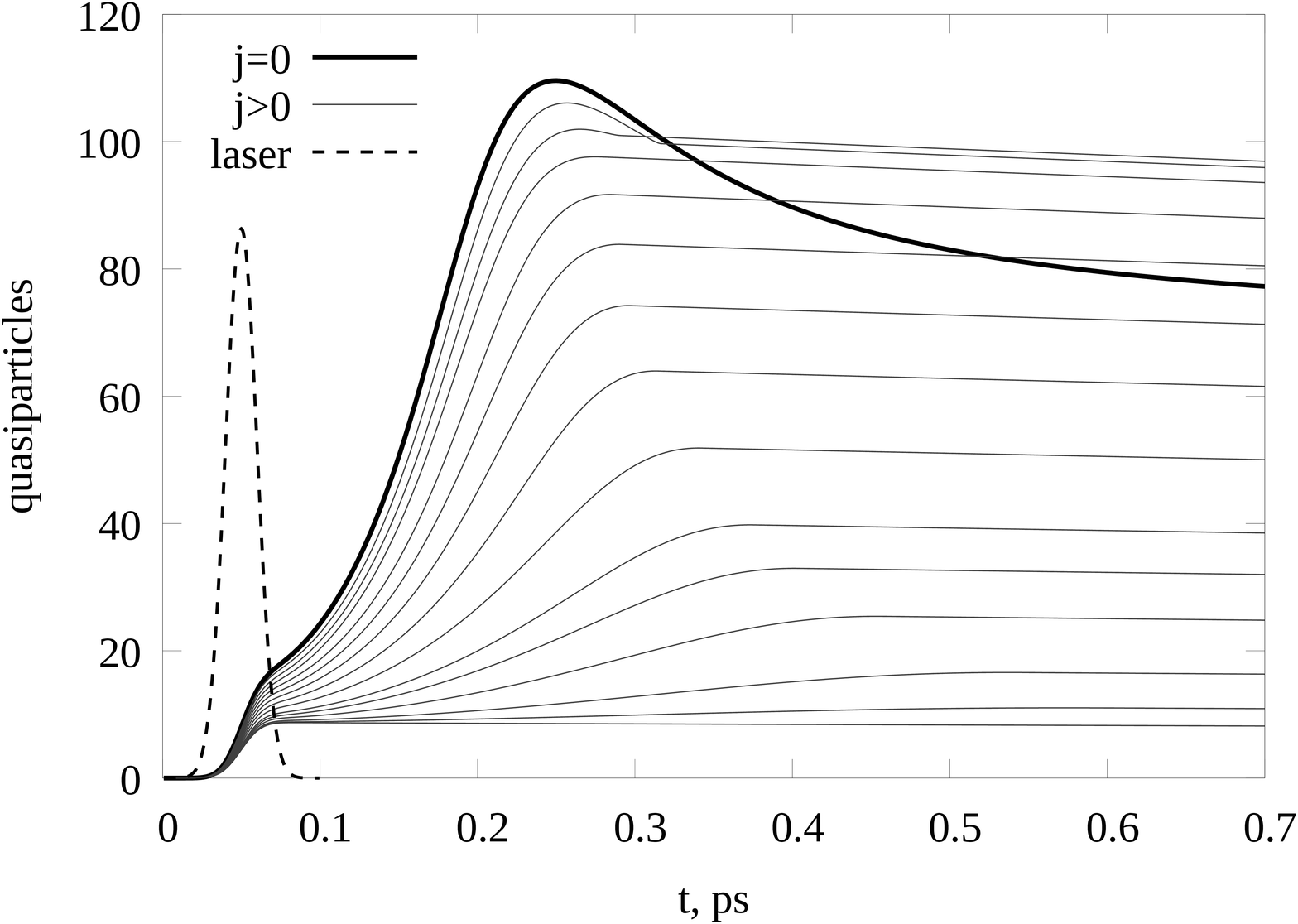}

(e) 
\end{center}
\end{minipage}

\begin{minipage}{0.5\linewidth}\begin{center}
\includegraphics[width=1.0\linewidth]{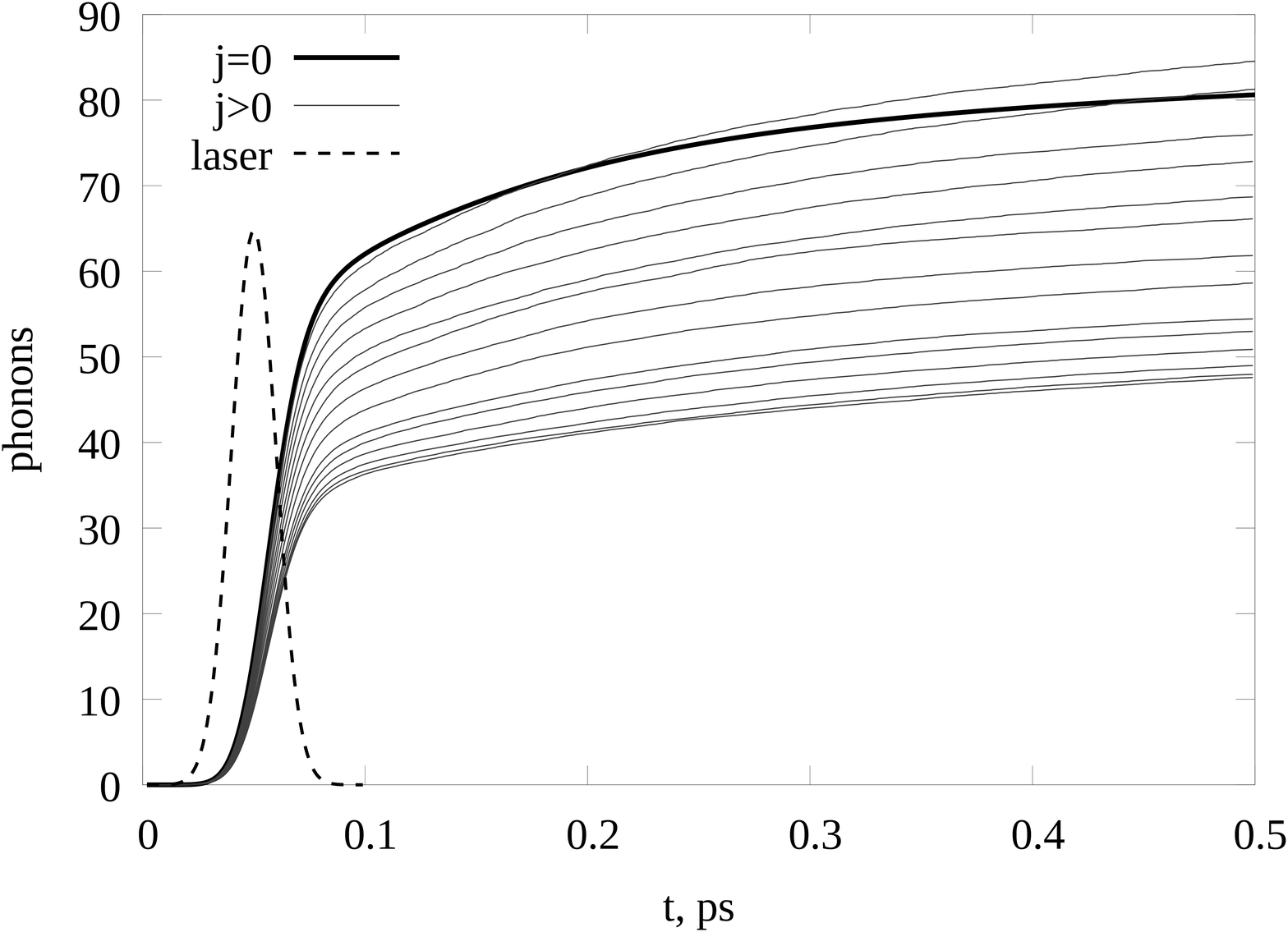}

(c) 
\end{center}
\end{minipage}
\begin{minipage}{0.5\linewidth}\begin{center}
\includegraphics[width=1.0\linewidth]{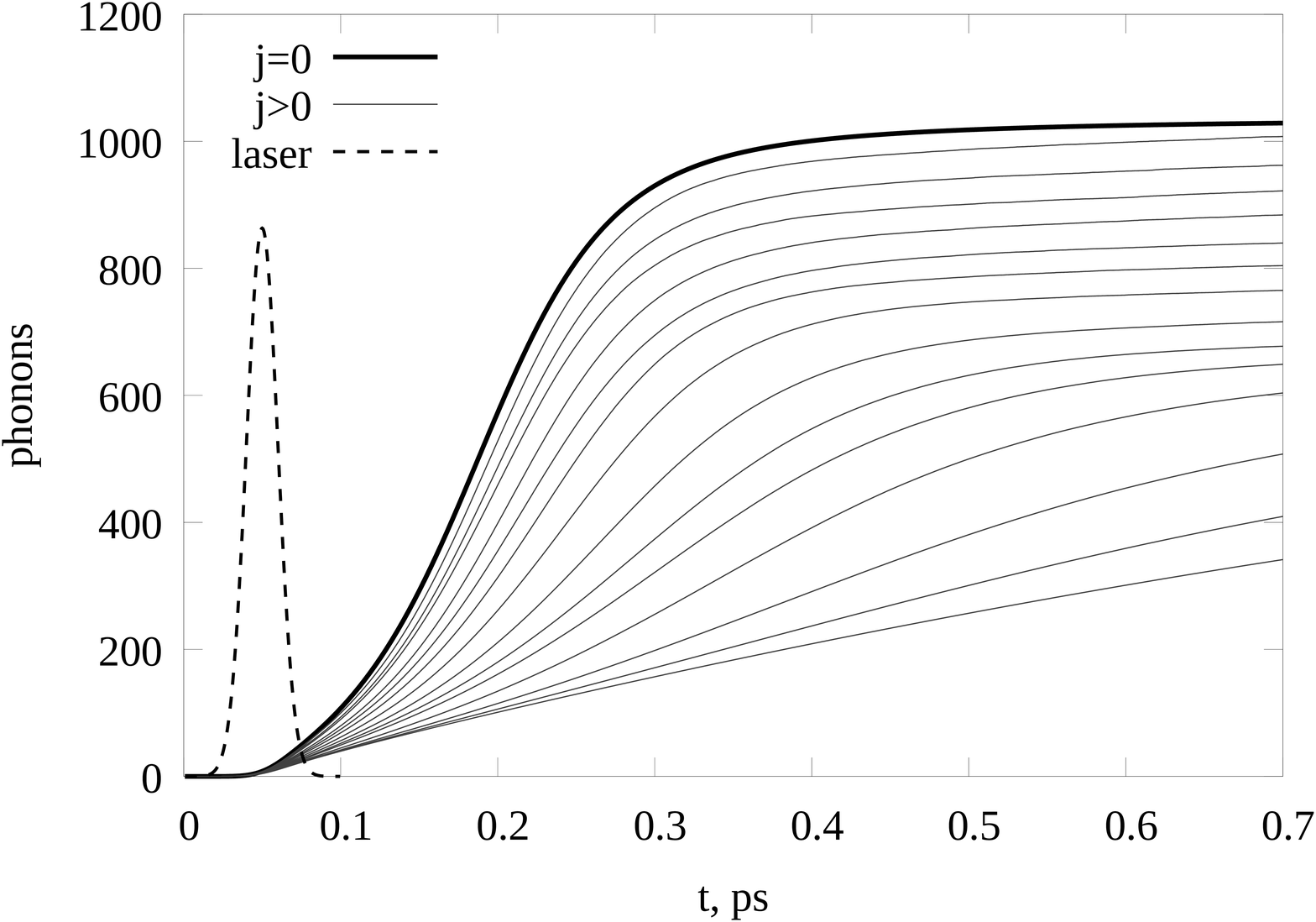}

(f) 
\end{center}
\end{minipage}
\begin{minipage}{0.5\linewidth}\begin{center}

\end{center}
\end{minipage}
\caption{
Time evolution of order parameter (a,d),
number of quasiparticles (b,e) and phonons (c,f) in the superconductor,
after the action of femtosecond pulse 
($\sigma$=10~fs, $I$=1.0),
calculated for several values of current density
$j$=0$\div$8.8$\cdot$10$^3$~A/cm$^2$,
at two values of radiation quantum energy
$\hbar \omega$=0.2~eV (a-c) and 3.0~eV (d-f).
}
\label{4_current_delta}
\end{figure}

\begin{figure}
\begin{center}
\includegraphics[width=0.5\linewidth]{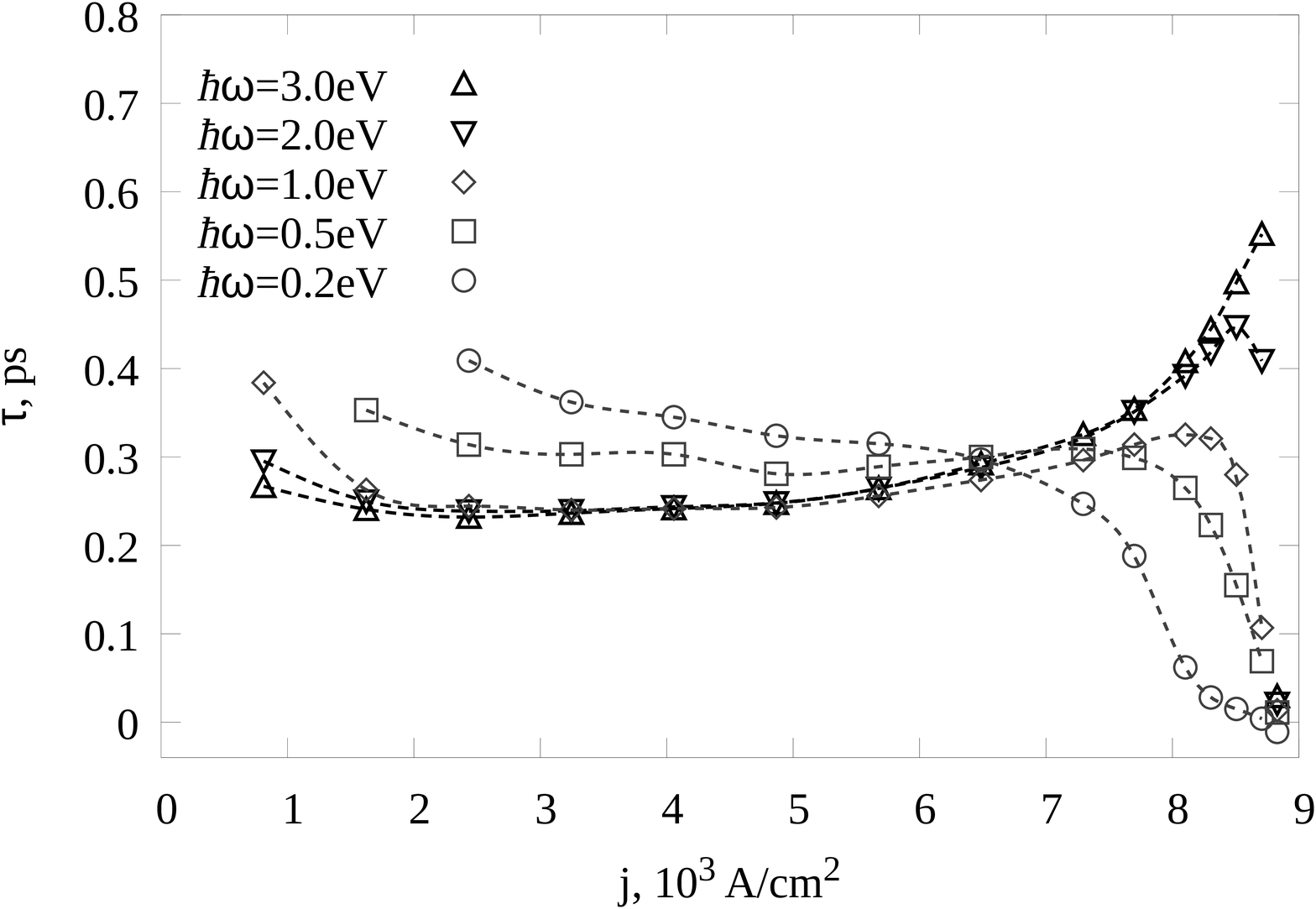}
\end{center}
\caption{
Time needed for the complete suppression of superconductivity
depending on the current density,
calculated for several energies of the radiation quantum.
Intensity $I$=1.0.
The missing points at low current densities
correspond to the cases 
when the complete suppression of order parameter was not achieved.
}
\label{5_current_vs_omega}
\end{figure}

In figures \ref{4_current_delta}(a)---\ref{4_current_delta}(f)
we show the process of superconductivity suppression
in the presense of nonzero current density,
calculated for two values of radiation quantum energy.
With increasing current density,
the initial value of the order parameter decreases.
This is the main factor accelerating 
the suppression of superconductivity
at low radiation frequency
(figures \ref{4_current_delta}(a)---\ref{4_current_delta}(c)).
At higher radiation frequency
(figures \ref{4_current_delta}(d)---\ref{4_current_delta}(f)), however,
the stage of the preliminary relaxation
of high-energy quasiparticles becomes important
(like in figure \ref{1_main_fs}).
In this case, the reduction of the generated quasiparticle density
due to the transport current can, instead, 
slow down the superconductivity suppression.

In figure \ref{5_current_vs_omega},
these features are demonstrated by the current dependence 
of the time needed for the complete suppression of superconductivity,
calculated for several quantum energies.
We conclude that 
for achieving the fast suppression of superconductivity with ultrashort laser pulse, 
a transport current can be beneficial
with the appropriate choice of the radiation frequency,
and conversely, for a given current density
the minimal switching time can be achieved with choosing the optimal laser wavelength.
For the parameters used in the simulation,
the longer wavelength for large current density
and the shorter wavelength for small current density are preferable.

\section{Conclusion}

In the framework of the problem 
of ulfrafast superconductivity switching with ultrashort laser pulse,
we study the relaxation of the nonequilibrium state
in the presence of transport current.
We use the qualitative theoretical model
describing the kinetics of nonequilibrium high-energy
quasiparticles and phonons 
accompanied by the relations of BCS theory
for the superconducting state.
The process of superconductivity suppression
is studied in details using numerical simulation.

While the high enough pulse energy can allow
to switch off the superconductivity completely
even during the pulse time,
it can cause harmful side effects
due to excess heat emission, 
and slow down the backwards switching.
In this work, we study the case of lower energy
with the delayed suppression of the superconducting order parameter
after the action of laser pulse.
The delay is caused by the finite time 
needed for the electron subsystem to lower the average energy due to relaxation,
as the order parameter is affected only by the low-energy quasiparticles.

The effect on the superconductivity suppression time
caused by the transport current was studied.
The results of numerical simulation
for various current densities
and laser radiation quantum energies demonstrate
that the suppression time is detemined by two main factors.
The first factor is the lowering of the initial value of order parameter
due to the presence of transport current, able to speed up the suppression.
The second factor is the delay of the first stage of relaxation,
depending on the radiation quantum energy.
It is found that the optimal current density 
needed to achieve the minimal switching time,
is substantially different for the cases of 
optical and infrared radiation ranges.

\section*{Acknowledgments}

The work is supported by the Russian Foundation
for Basic Research, Grant No.  17-29-10024.

\end{document}